\documentclass[aps,pra,reprint,amsmath,amssymb,superscriptaddress,showpacs,nobibnotes,twocolumn]{revtex4-1}
\usepackage{graphicx,SIunits}
\usepackage{bm}     
\usepackage{hyperref} 
\usepackage{dsfont}    
\usepackage{color}

\begin{document}

\title{Limits on manipulating conditional photon statistics via interference of weak lasers}

\author{Kang-Hee Hong}
\affiliation{Department of Physics, Pohang University of Science and Technology (POSTECH), Pohang, 37673, Republic of Korea}
\author{Ji-Sung Jung}
\affiliation{Center for Quantum Information, Korea Institute of Science and Technology (KIST), Seoul, 02792, Republic of Korea}
\affiliation{Department of Physics, Yonsei University, Seoul, 03722, Republic of Korea}
\author{Yong-Wook Cho}
\affiliation{Center for Quantum Information, Korea Institute of Science and Technology (KIST), Seoul, 02792, Republic of Korea}
\author{Sang-Wook Han}
\affiliation{Center for Quantum Information, Korea Institute of Science and Technology (KIST), Seoul, 02792, Republic of Korea}
\author{Sung Moon}
\affiliation{Center for Quantum Information, Korea Institute of Science and Technology (KIST), Seoul, 02792, Republic of Korea}
\author{Kyunghwan Oh}
\affiliation{Department of Physics, Yonsei University, Seoul, 03722, Republic of Korea}
\author{Yong-Su Kim}
\email{yong-su.kim@kist.re.kr}
\affiliation{Center for Quantum Information, Korea Institute of Science and Technology (KIST), Seoul, 02792, Republic of Korea}
\affiliation{Department of Nano-Materials Science and Engineering, Korea University of Science and Technology, Daejeon, 34113, Republic of Korea}
\author{Yoon-Ho Kim}
\email{yoonho72@gmail.com}
\affiliation{Department of Physics, Pohang University of Science and Technology (POSTECH), Pohang, 37673, Republic of Korea}

\date{\today}

\begin{abstract}
Photon anti-bunching, measured via the Hanbury-Brown--Twiss experiment, is one of the key signatures of quantum light and is tied to sub-Poissonian photon number statistics. Recently, it has been reported that photon anti-bunching or conditional sub-Poissonian photon number statistics can be obtained via  second-order interference of mutually incoherent weak lasers and heralding based on photon counting [Phys. Rev. A {\bf 92}, 033855 (2015); Opt. Express {\bf 24}, 19574 (2016); arXiv:1601.08161]. Here, we report theoretical analysis on the limits of manipulating conditional photon statistics via interference of weak lasers. It is shown that conditional photon number statistics can become super-Poissonian in such a scheme. We, however, demonstrate explicitly that it cannot become sub-Poissonian, i.e., photon anti-bunching cannot be obtained in such a scheme.  We  point out that incorrect results can be obtained if one does not properly account for seemingly negligible higher-order photon number expansions of the coherent state. 
\end{abstract}


\maketitle

\section{Introduction}
The normalized second-order correlation function $g^{(2)}(\tau)$ plays an important role in quantum optics by enabling to distinguish different kinds of light and is often measured  via the Hanbury-Brown-Twiss experiment \cite{hbt, glauber63}. Most importantly, it allows to distinguish different kinds of light. The chaotic light and the coherent light are associated with $g^{(2)}(0) \geq 1$ and they can be described well with the classical electromagnetic theory of light. Any light which exhibits $g^{(2)}(0)<1$, i.e., photon anti-bunching, is considered quantum as quantization of the electromagnetic field or the concept of photons is essential to describe its behaviors.  As the coherent state exhibits Poissonian photon number statistics, light which exhibits photon anti-bunching is tied to sub-Poissonian photon number statistics \cite{loudon}. There are indeed a wide variety of  quantum light, such as, the Fock states \cite{hong86,takeuchi04,baek08}, multi-photon entangled states \cite{yao12,jeong16,zhang16,wang16,chen17}, squeezed states \cite{breitenbach97,zhang15},  macroscopic superposition states \cite{sander92,ourj07}, etc., and applications of quantum light includes quantum communication \cite{gisin07}, quantum computing \cite{kok07}, quantum metrology \cite{giov11}, etc. 

A particular quantum state of light may be post-selected \cite{yao12,jeong16,zhang16,wang16,chen17,jeong13,lee14} or heralded \cite{takeuchi04,baek08,barz10, kim11,ra15,ra16}. In the heralding scheme, the statistical properties of the quantum state is conditioned by the heralding signal.  One of the earliest experiments on a localized single-photon state relied on the heralding signal from a single-photon detection event  of a two-photon state of spontaneous parametric down-conversion  \cite{hong86}. In this example, the heralding signal causes the conditional photon number statistics to be sub-Poissonian, exhibiting photon anti-bunching, whereas unheralded photon number statistics would be that of the chaotic light, i.e., super-Poissonian exhibiting photon bunching. In fact, quantum state heralding is a powerful tool in preparing a complex quantum state of light and  heralding schemes have been shown to generate a variety of non-classical light states, including various entangled states \cite{barz10, kim11,ra15,ra16}, photon added/subtracted states \cite{bellini}, etc. The idea of heralding has also been expanded to heralding quantum processes such as, quantum storage of light \cite{tanji09}, quantum gates and operations \cite{pittman03,xiang10,okamoto11}, etc.

Recently, it has been reported that photon anti-bunching or conditional sub-Poissonian photon number statistics can be obtained via  second-order interference of mutually incoherent weak lasers and heralding based on photon counting \cite{silva15,silva16,amaral16}. Even with weak lasers at the single-photon regime, it is well-known that linear optical elements do not change the photon number statistics and effects such as interference and phase randomization do not produce quantum light, although they may produce chaotic light which exhibits photon bunching or super-bunching \cite{cho10,hong12,pandey14,choi17}.  Then, the question becomes whether conventional single-photon detectors can indeed herald a non-classical light state exhibiting photon anti-bunching from the classical input light. In particular, what is the limit on manipulating conditional photon number statistics via interference of weak lasers and heralding based on photon counting?  

\begin{figure*}[t]
\centering
\includegraphics[width=6.5in]{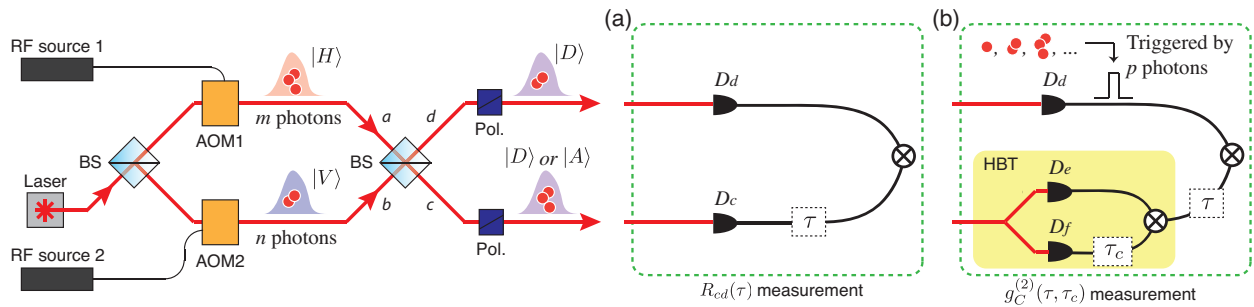}
\caption{The proposed experimental setup for manipulating photon statistics via second-order interference between two mutually incoherent weak lasers. AOM:  acousto-optic modulator, BS: beam splitter, Pol.:  polarizer, $D$: single-photon detector, HBT: Hanbury-Brown--Twiss interferometer. (a) This setup measures the second-order intensity cross-correlation $R_{cd}(\tau)$ between modes $c$ and $d$. (b) This setup measures the conditional second-order intensity autocorrelation $g_C^{(2)}(\tau,\tau_{c})$. Electronic delays $\tau$ and $\tau_{c}$ are used to explore various interference conditions.}
\label{fig1} 
\end{figure*}

In this paper, we report the theoretical analysis on the limits of manipulating conditional photon statistics via interference of weak lasers and heralding based on photon counting. It is shown that conditional photon number statistics can become super-Poissonian in such a scheme. We  demonstrate explicitly however that, contrary to Ref.~\cite{silva15,silva16,amaral16}, it cannot become sub-Poissonian, i.e., photon anti-bunching cannot be obtained in such a scheme. Theoretical and numerical analyses show that such incorrect results can be obtained if one does not properly account for seemingly negligible higher-order photon number expansions of the coherent state even at the single-photon regime.


\section{Experimental scheme}

Consider first the well-known Shih-Alley/Hong-Ou-Mandel experiment in which two single-photons impinge on a symmetric beam splitter via the two different input ports \cite{hom,shih}. As single-photons have no definite phase, no first-order interference is formed at the output of the beam splitter. If the two photons are distinguishable, they exit the beam splitter randomly. When coincidence counts are measured between the two detectors placed at each output port of the beam splitter, random coincidence events are measured. However, if the two single-photons are made to be indistinguishable, second-order quantum interference causes the photons to coalescence. As a result, the two photons are always found together at the same output port of the beam splitter, causing the change of the photon number statistics. In this case, null coincidence counts are measured due to the second-order quantum interference. The  visibility, defined as the random coincidence counts subtracted by the coincidence counts due to quantum interference normalized to the random coincidence counts,  in this case can reach the maximum value of one.

The experimental setup to manipulate conditional photon statistics via second-order interference of mutually incoherent weak lasers and heralding based on photon counting is inspired by the above mentioned Shih-Alley/Hong-Ou-Mandel experiment. Instead of single-photon states at the input of a symmetric beam splitter, we have two mutually incoherent weak lasers at the two input ports, see Fig.~\ref{fig1}. The setup is similar to the experiments in Ref.~\cite{choi17,yongsu,yongsu2,silva15,silva16,amaral16} but we  make use of the additional polarization degree of freedom  to more clearly present the idea. Two beams of lasers $a$ and $b$ are prepared by splitting a laser beam with a beam splitter (BS). The mutual phase coherence between the two beams $a$ and $b$ is removed by using the acousto-optic modulators (AOM1 and AOM2) that are driven by independent RF sources~\cite{yongsu,yongsu2}. This is essential to ensure that there is no first-order interference between the two beams $a$ and $b$ as we overlap the two beams at the second BS. This configuration corresponds to the Shih-Alley/Hong-Ou-Mandel experiment but with two classical light beams \cite{hom,shih,yongsu,yongsu2}. 

The second-order intensity correlation between the two output ports of the BS, $c$ and $d$, can be manipulated by changing the interference condition. Figure~\ref{fig1}(a) shows a typical experimental setup to measure the second-order intensity correlation $R_{cd}(\tau)$ between the modes $c$ and $d$. Here, $\tau$ denotes the time delay between two single-photon detection events at $D_c$ and $D_d$. The interference condition can be easily changed by the polarization states of beams. In order to implement various interference conditions, the polarization states of the laser beams at $a$ and $b$ are set to be  horizontal ($|H\rangle_a$) and vertical ($|V\rangle_b$), respectively. Then, the interference condition can be manipulated by changing the combinations of the polarization projection bases at the output modes $c$ and $d$ by using the polarizers (Pol). When the projection  basis $\{|D\rangle_c,|D\rangle_d\}$, where $|D\rangle=\frac{1}{\sqrt{2}}(|H\rangle + |V\rangle)$, is chosen for the polarizers, the second-order intensity correlation $R_{cd}(0)$ reaches the minimum value. For the projection basis $\{|A\rangle_c,|D\rangle_d\}$, where $|A\rangle=\frac{1}{\sqrt{2}}(|H\rangle - |V\rangle)$, the second-order intensity correlation $R_{cd}(0)$ reaches the maximum value.  We thus consider these bases in our analysis as they offer the theoretical maximum visibility of the second-order interference \cite{choi17,yongsu,yongsu2}. 

Since the value of the second-order intensity correlation $R_{cd}(\tau>\tau_{\rm coh})=1$, where $\tau_{\rm coh}$ is the coherence time of the input light, if the input light is two mutually incoherent lasers, $R_{cd}(0)=0.5$ for the  projection  basis $\{|D\rangle_c,|D\rangle_d\}$. Thus, the visibility of the  second-order classical interference in the Shih-Alley/Hong-Ou-Mandel setup is  limited by $V\le0.5$ \cite{choi17,yongsu,yongsu2}. This result nevertheless signals that photons are weakly bunched. On the other hand, if the measurement basis  $\{|A\rangle_c,|D\rangle_d\}$ is chosen,  $R_{cd}(0) =1.5$ indicating that the photons tend to distribute themselves in different spatial modes.  This simple argument allows us to ask what the limit would be for manipulating conditional photon statistics in a particular output mode $c$ by using photon counting at the other output mode $d$ as the heralding signal. In particular, we are interested in the conditional second-order intensity correlation function for mode $c$, $g^{(2)}_C(\tau,\tau_{c})$, heralded by the photon counting signal at mode $d$. The relevant experimental setup to measure $g^{(2)}_C(\tau,\tau_{c})$ is shown in Fig.~\ref{fig1}(b). Conditioned on the photon counting event at mode $d$, we use the Hanbury-Brown--Twiss setup to measure $g^{(2)}_C(\tau,\tau_{c})$, the conditional second-order intensity correlation function at mode $c$.

\section{Conditional photon statistics}

In this section, we show the state evolution following the experimental setup in Fig.~\ref{fig1}. The input to the first BS is a weak laser and its quantum state can be written in the Fock basis as,
\begin{equation}
|\Psi\rangle_{{\rm in}}=e^{-\frac{\alpha}{2}} \sum_{\lambda=0}^{\infty} \frac{\alpha^{\frac{\lambda}{2}}}{\sqrt{\lambda!}} |\lambda\rangle,
\label{input}
\end{equation}
where $|\lambda\rangle$ denotes the $\lambda$-photon Fock state. The  symmetric 50/50 beam splitter splits the incoming beam into two, each beam having the mean photon number corresponding to  $\alpha/2$. The quantum state of light at the output of the first BS is thus given by
\begin{eqnarray}
|\Psi\rangle &=& e^{-\frac{\alpha}{4}} \sum_{m=0}^{\infty} \frac{(\frac{\alpha}{2})^{\frac{m}{2}}}{\sqrt{m!}} | m\rangle_a \otimes e^{-\frac{\alpha}{4}} \sum_{n=0}^{\infty} \frac{(\frac{\alpha}{2})^{\frac{n}{2}}}{\sqrt{n!}} | n\rangle_b\nonumber\\
&=& e^{-\frac{\alpha}{2}} \sum_{m=0}^{\infty} \sum_{n=0}^{\infty} \frac{(\frac{\alpha}{2})^{\frac{n+m}{2}}}{m! n! } (\hat a^\dagger)^m (\hat b^\dagger)^n| 0\rangle
\end{eqnarray}
where $|m\rangle_a$ and  $|n\rangle_b$ represent the $m$ and $n$ photon Fock states, respectively. The photon creation  operators at modes $a$ and $b$ are denoted as $\hat a^\dag$ and $\hat b^\dag$, respectively. 

At each output of the first BS, an acousto-optic modulator (AOM) is placed. The AOMs are driven by independent and unsynchronized RF sources, making the incoming light beams mutually incoherent.  Therefore, after the AOMs, the state of each frequency mode can be represented as 
\begin{eqnarray}
|\Psi(\omega_a,\omega_b)\rangle &=& e^{-\frac{\alpha}{2}} \sum_{m,n=0}^{\infty} \frac{(\frac{\alpha}{2})^{\frac{n+m}{2}}}{m! n! }\nonumber\\
&\times&
\left\{e^{i\gamma_{\omega_a}}\hat a^\dagger(\omega_a)\right\}^m \left\{e^{i\gamma_{\omega_b}}\hat b^\dagger(\omega_b)\right\}^n| 0\rangle,
\label{Psi}
\end{eqnarray}
where $\omega_{a} (\omega_{b})$ and $\gamma_{\omega_a} (\gamma_{\omega_b})$ are the frequency mode and the phase given by the AOM1 (AOM2) at mode $a~(b)$, respectively. The state of light beams at the input mode $a$ and $b$ to the second BS thus can be written as,
\begin{equation}
\rho=\int d\omega_a d\omega_b \mathcal{F}(\omega_a)\mathcal{F}(\omega_b)|\Psi(\omega_a,\omega_b)\rangle\langle\Psi(\omega_a,\omega_b)|,
\label{rho}
\end{equation}
where $\mathcal{F}(\omega_a)$ and $\mathcal{F}(\omega_b)$ are the frequency spectra of the light given by AOM1 and AOM2, respectively. The quantum description of the light beams at the input modes $a$ and $b$ to the second BS can then be fully written as,
\begin{eqnarray}
\rho &=& e^{-\alpha}\int d\omega_a \mathcal{F}(\omega_a)\sum_{m=0}^{\infty} \frac{(\frac{|\alpha|}{2})^{m}}{m!}|m\rangle_a\langle m|\nonumber\\
 &&~~ \times\int d\omega_b \mathcal{F}(\omega_b)\sum_{n=0}^{\infty} \frac{(\frac{|\alpha|}{2})^{n}}{n!}|n\rangle_b\langle n|.
\label{eq_afint}
\end{eqnarray}
Note that the photons in mode $a$ and $b$ are horizontally and vertically  polarized, respectively. Since the BS input modes $a$ and $b$ are related to the output modes $c$ and $d$ according to the following relation, 
\begin{eqnarray}
\hat a^\dagger_H(\omega_a)&\to&\frac{1}{\sqrt{2}}\hat c^\dagger_H(\omega_a)+\frac{i}{\sqrt{2}}\hat d^\dagger_H(\omega_a),\nonumber\\
\hat b^\dagger_V(\omega_b)&\to&\frac{i}{\sqrt{2}}\hat c^\dagger_V(\omega_b)+\frac{1}{\sqrt{2}}\hat d^\dagger_V(\omega_b),
\label{bs_transform}
\end{eqnarray}
the state of light at the output modes $c$ and $d$ of the second BS is given by,
\begin{eqnarray}
\rho_{cd} &=& e^{-\alpha}\int d\omega_ad\omega_b \mathcal{F}(\omega_a)\mathcal{F}(\omega_b)\sum_{m,n}\frac{(\frac{|\alpha|}{2})^{n+m}}{(m! n!)^2}\left(\frac{1}{2}\right)^{m+n}\nonumber\\
&&\times\left\{\hat c_H^\dagger(\omega_a)+i \hat d^\dagger_H(\omega_a)\right\}^m\nonumber\\
&&\times\left\{i \hat c_V^\dagger(\omega_b)+\hat d^\dagger_V(\omega_b)\right\}^n |0\rangle\langle 0|\{C.C.\},
\end{eqnarray}
where $\{C.C.\}$ denotes the complex conjugate.


Let us first consider the second-order intensity cross correlation, $R_{cd}(\tau)$, between modes $c$ and $d$, defined as
\begin{equation}
R_{cd}(\tau)=\frac{\text{Tr}\left[\rho_{cd} E^{(-)}_c(\tau) E^{(-)}_d(0) E^{(+)}_d(0) E^{(+)}_c(\tau)\right]}{\text{Tr}\left[\rho_{cd} E^{(-)}_c(\tau)E^{(+)}_c(\tau)\right] \text{Tr}\left[\rho_{cd} E^{(-)}_d(0) E^{(+)}_d(0)\right]},
\label{R_cd}
\end{equation}
where $E^{(+)}_c(t)=\frac{1}{\sqrt{2\pi}}\int  \hat c(\omega)e^{-i\omega t}d\omega$ is the field operator at detector $D_c$  and $\hat c(\omega)$ is the photon annihilation operator in mode $c$. As mentioned in section II, polarization projection precedes  photon detection and to ensure maximum interference visibility, the polarizer angles are set at $|D\rangle_c$ or $|A\rangle_c$. The polarizer is accounted for in the above equation by defining $\hat c(\omega)\equiv\frac{1}{\sqrt{2}}(\hat c_H(\omega) + \hat c_V(\omega))$ for the polarizer angle setting at $|D\rangle_c$. For polarization projection at $|A\rangle_c$, $\hat c(\omega)\equiv\frac{1}{\sqrt{2}}(\hat c_H(\omega) - \hat c_V(\omega))$. $E^{(+)}_d(t)$ and $\hat d(\omega)$ are defined similarly.  In our analysis, we consider the polarization projection bases $\{|D\rangle_c$, $|D\rangle_d\}$ and $\{|A\rangle_c$, $|D\rangle_d\}$ as they offer the theoretical maximum visibility of the second-order interference \cite{choi17,yongsu,yongsu2}. Note that  Eq.~(\ref{R_cd}) can be measured  by using the experimental setup shown in Fig.~\ref{fig1}(a). 

Let us now consider conditional photon statistics at mode $c$, heralded by photon detection event at mode $d$. The relevant experimental setup is shown in Fig.~\ref{fig1}(b). To investigate the conditional second-order correlation $g_C^{(2)}(\tau,\tau_{c})$, an Hanbury-Brown--Twiss setup is placed in mode $c$ and is triggered by the photon counting signal from detector $D_d$. Assuming the single-photon detectors cannot resolve the number of photons, the second-order correlation at mode $c$ heralded by photon detection at mode $d$ can be written  as,
\begin{equation}
g^{(2)}_{C}(\tau,\tau_{c})=\frac{\sum_{p=1}^{\infty} I^{(p)}_d \sum_{p=1}^{\infty} I_{def}^{(p)}(\tau,\tau_{c})}{\sum_{p=1}^{\infty} I_{de}^{(p)}(\tau) \sum_{p=1}^{\infty} I_{df}^{(p)}(\tau+\tau_{c})}.
\label{eq_cg2}
\end{equation}
Here, $I_d^{(p)}$ is the heralding probability caused by a $p$-photon detection at detector $D_d$. Since the photon counting detectors are not photon number resolving, a heralding signal may be caused either by a  single-photon or by multiple photons. The heralding probability is  given as,
\begin{equation}
I_d^{(p)}=\text{Tr}\left[\rho_{cd}\{E_d^{(-)}(0)\}^p\{E_d^{(+)}(0)\}^p\right].
\label{I_1}
\end{equation}
The two-fold coincidence probability $I_{de}^{(p)}$ ($I_{df}^{(p)}$) describes the coincidence events between the detectors $D_d$ and $D_e$  ($D_d$ and $D_f$), assuming that $p$-photon detection event occurs at detector $D_d$. $I_{de}^{(p)}$ and $I_{df}^{(p)}$ are given by, 
\begin{eqnarray}
I_{de}^{(p)}(\tau)&=&\text{Tr}\left[\rho_{cd}\{E_d^{(-)}(0)\}^p E_e^{(-)}(\tau) E_e^{(+)}(\tau) \{E_d^{(+)}(0)\}^p\right],\nonumber\\
I_{df}^{(p)}(\tau)&=&\text{Tr}\left[\rho_{cd}\{E_d^{(-)}(0)\}^p E_f^{(-)}(\tau) E_f^{(+)}(\tau) \{E_d^{(+)}(0)\}^p\right].\nonumber
\label{I_de}
\end{eqnarray}
Here, $E^{(+)}_e(t)=\frac{1}{\sqrt{2\pi}}\int  \hat e(\omega)e^{-i\omega t}d\omega $ and  $E^{(+)}_f(t)=\frac{1}{\sqrt{2\pi}}\int  \hat f(\omega)e^{-i\omega t}d\omega $ are the field operators at detector $D_e$ and $D_f$, respectively, with the annihilation operators $\hat e(\omega)=\frac{1}{\sqrt{2}}\hat c(\omega)$ and $\hat f(\omega)=\frac{i}{\sqrt{2}}\hat c(\omega)$. 

The conditional coincidence count rate between $D_e$ and $D_f$, heralded by the photon detection event at $D_d$, is given by the triple coincidence probability,
\begin{eqnarray}
I_{def}^{(p)}(\tau,\tau_{c})&&=\text{Tr}\Bigl[\rho_{cd}\{E_d^{(-)}(0)\}^p E_e^{(-)}(\tau) E_f^{(-)}(\tau+\tau_{c})\nonumber\\
&&~~~\times E_f^{(+)}(\tau+\tau_{c}) E_e^{(+)}(\tau) \{E_d^{(+)}(0)\}^p\Bigr].
\label{I_def}
\end{eqnarray}

\begin{figure}[t]
\centering
\includegraphics[width=3.2in]{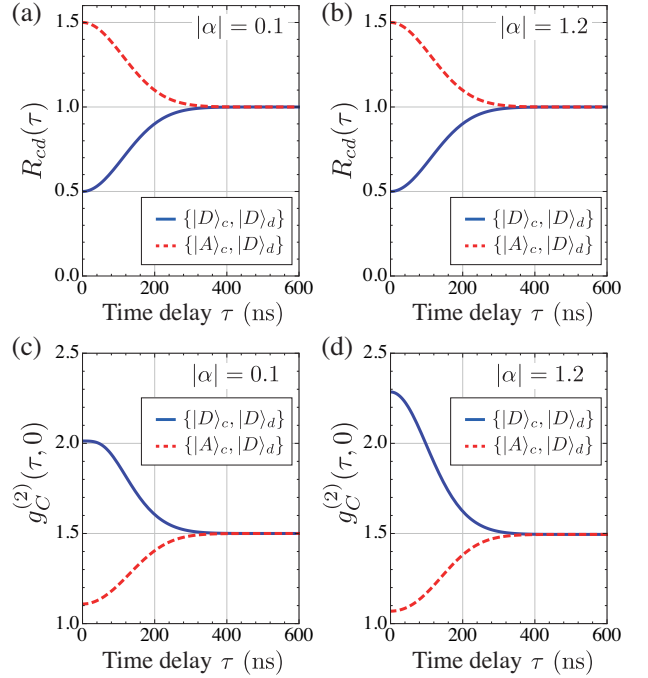}
\caption{The intensity cross-correlation $R_{cd}(\tau)$ and the conditional second-order  correlation $g_C^{(2)}(\tau,0)$ for $|\alpha|=0.1$ and $|\alpha|=1.2$. Both $R_{cd}(\tau)$ and $g_C^{(2)}(\tau,0)$ are truncated at $p=10$.  The  coherence time $\tau_{\rm coh} \approx 260$ ns. The blue solid lines and red dashed lines correspond to projection measurement bases $\{|D\rangle_c,|D\rangle_d\}$, and $\{|A\rangle_c,|D\rangle_d\}$, respectively. Note that $g_C^{(2)}(\tau,0)$ is never below 1, meaning that the heralded photon states always remain classical.}
\label{result_1}
\end{figure}

\section{Analysis}

Even at the single-photon regime of $|\alpha|^2$, the fact that the laser follows the Poisson photon number statistics makes the analytical analysis difficult. Nevertheless, the conditional second-order correlation $g_C^{(2)}(\tau,\tau_{c})$ in Eq.~(\ref{eq_cg2}) can be obtained analytically if we  approximate the initial input state of Eq.~(\ref{input}) up to a finite photon number $p$. In the following analysis, we consider the analytic forms of $g_C^{(2)}(\tau,\tau_c)$ truncated at the $p$-photon Fock state term. We  consider the case of $\tau_c=0$   because this is where the quantum nature of the heralded photon states can be best tested. For instance,  the conditional second order correlation function truncated at $p=3$ for an arbitrary $\alpha$, is given as, 
\begin{equation}
g^{(2)}_{C,p=3}(\tau,0)=\mp\frac{\left(129 \alpha ^2+176 \alpha +128\right) \left(2 e^{\frac{\sigma ^2 \tau ^2}{2}}\mp3 e^{\sigma ^2 \tau ^2}\right)}{4 \left((11 \alpha +8) e^{\frac{\sigma ^2 \tau ^2}{2}}\mp6 \alpha \mp4\right)^2},\nonumber
\end{equation}
where $\sigma$ denotes the bandwidth of the Gaussian-shaped frequency spectra of $\mathcal{F}(\omega_a)$ and $\mathcal{F}(\omega_b)$ in Eq.~(7). Here, the $\mp$ sign is related to the polarization projection measurement  $\{|D\rangle_c$, $|D\rangle_d\}$ and $\{|A\rangle_c$, $|D\rangle_d\}$.   Note that, while truncating at a low photon number $p$ makes the analytic forms simpler, it could significantly alter the conditional second order correlation $g_C^{(2)}(\tau,0)$. The analytic forms of $g_C^{(2)}(\tau,0)$ up to $p=10$ are given in Appendix. 

Figure 2 shows the numerical simulation of the intensity cross-correlation $R_{cd}(\tau)$ and the conditional second-order  correlation $g_C^{(2)}(\tau,0)$ for weak coherent states with $|\alpha|=0.1$ and $|\alpha|=1.2$. To account for sufficiently large Fock state contributions, truncation is made at $p=10$. Also, we assumed that the frequency spectra $\mathcal{F}(\omega_a)$ and $\mathcal{F}(\omega_b)$ in Eq.~(7) to be Gaussian with the bandwidth of $15/2\pi$ MHz in full width at half maximum (FWHM), corresponding to the coherence time $\tau_{\rm coh} \approx 260$ ns. The blue solid lines and red dashed lines correspond to projection measurement bases $\{|D\rangle_c,|D\rangle_d\}$, and $\{|A\rangle_c,|D\rangle_d\}$, respectively. The $R_{cd}(\tau)$ plots  in Fig.~2(a) and Fig.~2(b) show the typical Shih-Alley/Hong-Ou-Mandel like peaks and dips with the limited visibility of $V=1/2$~\cite{hom,shih,yongsu,yongsu2}. The conditional second order correlation $g_C^{(2)}(\tau,0)$ is shown in Fig.~2(c) and Fig.~2(d). While it is evident that the conditional photon statistics can be manipulated, even becoming super-bunched, i.e., $g_C^{(2)}(\tau,0)>2$ as in Fig.~2(d), $g_C^{(2)}(\tau,0)$ is never below 1, meaning that the heralded photon states always remain classical.

\begin{figure}[t]
\centering
\includegraphics[width=2.6in]{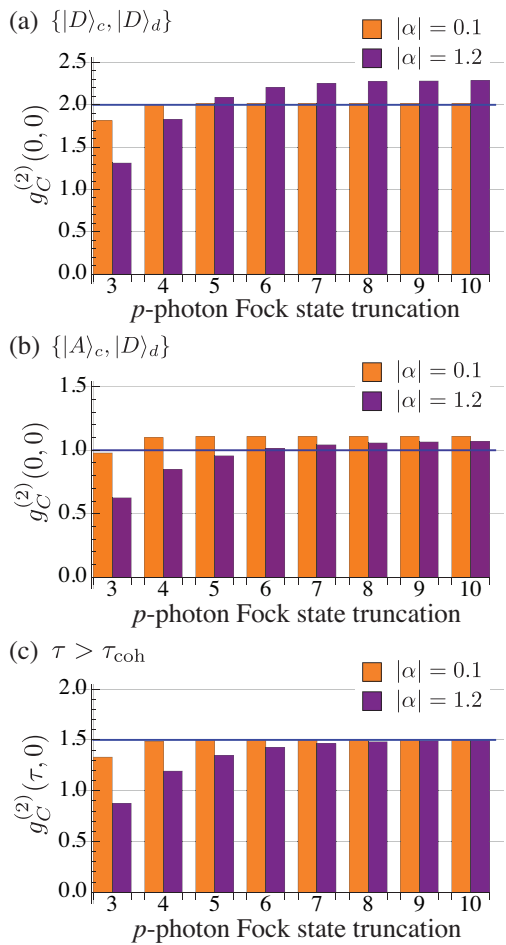}
\caption{The conditional second order correlation $g_C^{(2)}(0,0)$ for $|\alpha|=0.1$ and $|\alpha|=1.2$ under different  $p$-photon Fock state truncation. For polarization projection (a) $\{|D\rangle_c,|D\rangle_d\}$ and  (b) $\{|A\rangle_c,|D\rangle_d\}$. (c)  $g_C^{(2)}(\tau,0)$ with $\tau=500$ ns, i.e., $\tau > \tau_{\rm coh}$.  Even for weak coherent state at the single-photon regime,  $|\alpha|=0.1$, asymptotic behaviors are observed at relatively large $p=4$.  For  $|\alpha|=1.2$, asymptotic behaviors are not reached until $p=9$, meaning that truncation below $p=9$ would result incorrect results.  If Fock state truncation is made before reaching the asymptotic value corresponding to a particular $\alpha$, it looks as though conditional photon anti-bunching were possible. }
\label{result_2} 
\end{figure}

In Fig.~\ref{result_2}, we show the conditional second order correlation $g_C^{(2)}(\tau,\tau_c)$ for $|\alpha|=0.1$ and $|\alpha|=1.2$ under different $p$-photon Fock state truncation. By looking at the asymptotic behaviors of the second order correlation function, we can figure out whether truncation at the particular $p$-photon Fock state can be justified. First, the   $g_C^{(2)}(\tau,\tau_c)$ values are calculated for the condition $\tau=0$ and $\tau_c=0$, see Fig.~\ref{result_2}(a) and Fig.~\ref{result_2}(b). As noticed in Fig.~\ref{result_1}, different polarization projections $\{|D\rangle_c$, $|D\rangle_d\}$ and $\{|A\rangle_c$, $|D\rangle_d\}$ result in different conditional photon number statistics. 

Now, when $\tau$ is  larger than the coherence time of the light, $\tau>\tau_{\rm coh} \approx 260$ ns, the conditional second order correlations are the same regardless of the polarization projection choices. Figure~\ref{result_2}(c) shows  $g_C^{(2)}(\tau,0)$ at $\tau=500$ ns. The conditional second order correlation  $g_C^{(2)}(\tau>\tau_{\rm coh},0)$ starts out showing photon anti-bunching when the Fock state contributions are truncated at a low photon number $p$. However, as more and more  $p$-photon Fock state components are taken into consideration, it reaches  the asymptotic value of 1.5, which corresponds to the case when there is no heralding signal~\cite{pandey14}. 

What we find in Fig.~\ref{result_2} is that even for weak coherent state at the single-photon regime,  $|\alpha|=0.1$, asymptotic behaviors are observed at relatively large $p=4$.  For  $|\alpha|=1.2$, asymptotic behaviors are not reached until $p=9$, meaning that truncation below $p=9$ would result incorrect results. If Fock state truncation is made before reaching the asymptotic value corresponding to a particular $\alpha$, it looks as though conditional photon anti-bunching were possible \cite{silva15,silva16,amaral16}.  Such conditional photon anti-bunching from classical light, however, is purely due to improper handling of Fock state truncation. It is necessary to properly account for even seemingly negligible higher-order Fock state components of the coherent state.

\begin{figure*}[t]
\centering
\includegraphics[width=5.6in]{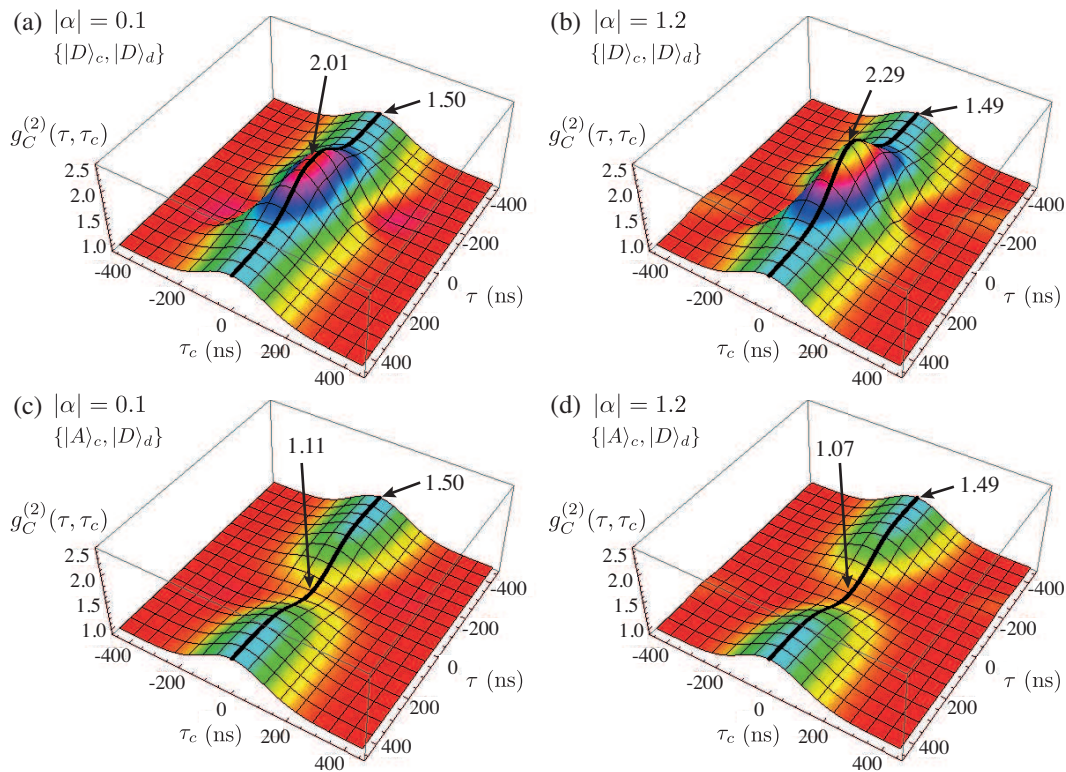}
\caption{The conditional second order correlation $g^{(2)}_C(\tau,\tau_{c})$, truncated at the $p=10$ photon Fock state, as functions of both $\tau$ and $\tau_{c}$. The black solid lines correspond to the case of $\tau_{c} = 0$, presented in Fig.~\ref{result_1}. It is clear that photon antibunching cannot be achieved by heralding if the input light is classical.}
\label{result_3} 
\end{figure*}

Figure~\ref{result_3} shows the conditional second order correlation $g^{(2)}_C(\tau,\tau_{c})$, truncated at the $p=10$ photon Fock state, as functions of both $\tau$ and $\tau_{c}$. The results with the $\{|D\rangle_c$, $|D\rangle_d\}$ projection are presented in Fig.~\ref{result_3}(a) and (b), while Fig.~\ref{result_3}(c) and (d) show the results with the $\{|A\rangle_c$, $|D\rangle_d\}$ projection. The black solid line in each plot corresponds to the case of $\tau_{c}=0$, which is depicted in Fig.~\ref{result_1}. It is clear that photon antibunching cannot be achieved by heralding if the input light is classical.

We also note that, when $\tau>\tau_{\rm coh}$, Figs.~\ref{result_1}, \ref{result_2}, \ref{result_3} show that  $g_C^{(2)}(\tau,0)=1.5$ which is larger than the value for the coherent state. This increased value of the second order correlation is due to the fact that the light in mode $c$ is the result of interference between two mutually incoherent laser beams \cite{pandey14}.

\section{Conclusion}

Generation of nonclassical light states is important in quantum optics and inexpensive and simple methods of generating various nonclassical light states would significantly advance the state of the art in experimental quantum information research. In a recent attempt to tackle such a challenging problem, it has been reported that photon anti-bunching or conditional sub-Poissonian photon number statistics can be obtained via second-order interference of mutually incoherent weak lasers and heralding based on photon counting~\cite{silva15,silva16,amaral16}. Here, we have carried out detailed theoretical and numerical analyses on the limit of manipulating conditional photon number statistics via interference of weak lasers and heralding based on photon counting. We find that conditional photon number statistics can become super-Poissonian in such a scheme. We demonstrate explicitly however that it cannot become sub-Poissonian, i.e., photon anti-bunching cannot be obtained in such a scheme. Theoretical and numerical analyses show that such incorrect results can be obtained if one does not properly account for seemingly negligible higher-order photon number expansions of the coherent state even at the single-photon regime. 

Additionally, our work shows that a light beam having super-Poissonian photon number statistics and photon super-bunching may be easily generated by interfering multiple mutually incoherent laser beams. Such a simple scheme for generating super-bunched light may be of use in optical super-resolution  and ghost imaging/interference experiments \cite{cao08, zhou10}.


\section*{Acknowledgement}

This work was supported by the National Research Foundation (Grant No.~2016R1A2A1A05005202 and No.~2016R1A4A1008978), the ICT R$\&$D program of MSIP/IITP (B0101-16-1355), and the KIST institutional program (Project No. 2E26680-16-P024, 2E27231, 2V05340).


\clearpage

\appendix*
\onecolumngrid
\section{Solutions for $g^{(2)}_C(\tau,\tau_{c})$}

Here, we show the analytical forms of $g^{(2)}_{C}(\tau,0)$ in Eq.~(\ref{eq_cg2}), truncated at the $p$-photon Fock state, for the projection basis  $\{|D\rangle_c,|D\rangle_d\}$. The cases up to $p=10$ are shown here. 
\begin{eqnarray}
g^{(2)}_{C,p=4}(\tau,0)=\frac{2 \left(5047 \alpha ^3+9288 \alpha ^2+12672 \alpha +9216\right) e^{-2 \sigma ^2 \tau ^2} \left(\alpha  e^{\sigma ^2 \tau ^2}-16 (3 \alpha
   +2) e^{\frac{5 \sigma ^2 \tau ^2}{2}}+6 (11 \alpha +8) e^{3 \sigma ^2 \tau ^2}\right)}{9 \left(\left(258 \alpha ^2+352 \alpha +256\right)
   e^{\frac{\sigma ^2 \tau ^2}{2}}-155 \alpha ^2-192 \alpha -128\right)^2}.\nonumber
\end{eqnarray}

\begin{eqnarray}
g^{(2)}_{C,p=5}(\tau,0)=\frac{9}{128} && \left(1221863 \alpha ^4+2584064 \alpha ^3+4755456 \alpha ^2+6488064 \alpha +4718592\right)e^{-\sigma ^2 \tau ^2}\nonumber\\
&&\times \Bigl[-2 \left(155 \alpha ^2+192 \alpha +128\right) e^{\frac{3 \sigma ^2 \tau
   ^2}{2}}+\left(387 \alpha ^2+528 \alpha +384\right) e^{2 \sigma ^2 \tau ^2}+2 \alpha  (7 \alpha
   +4)\Bigr]\nonumber\\
 &&  \Big/ \Bigl[3347 \alpha ^3+5580 \alpha ^2-\left(5047 \alpha ^3+9288 \alpha ^2+12672 \alpha
   +9216\right) e^{\frac{\sigma ^2 \tau ^2}{2}}+6912 \alpha +4608\Bigr]^2.\nonumber
\end{eqnarray}
   
\begin{eqnarray}   
   g^{(2)}_{C,p=6}(\tau,0)=\frac{64}{25} &&\left(110286921 \alpha ^5+244372600 \alpha ^4+516812800 \alpha ^3+951091200 \alpha^2 \right.\left. +1297612800 \alpha +943718400\right)e^{-\sigma ^2 \tau ^2} \nonumber\\
   &&\times \Bigl[9 \alpha  \left(213 \alpha ^2+224
   \alpha +128\right)-8 \left(3347 \alpha ^3+5580 \alpha ^2+6912 \alpha +4608\right) e^{\frac{3 \sigma^2 \tau ^2}{2}}\nonumber\\
   &&+6 \left(5047 \alpha ^3+9288 \alpha ^2+12672 \alpha +9216\right) e^{2 \sigma ^2 \tau^2}\Bigr]\nonumber\\
   &&\Big/ \Bigl[1772967 \alpha ^4+3427328 \alpha ^3+5713920 \alpha ^2+7077888 \alpha +4718592\nonumber\\
   &&-2 \left(1221863 \alpha
   ^4+2584064 \alpha ^3+4755456 \alpha ^2+6488064 \alpha +4718592\right) e^{\frac{\sigma ^2 \tau
   ^2}{2}}\Bigr]^2.\nonumber
\end{eqnarray}

\begin{eqnarray}
   g^{(2)}_{C,p=7}(\tau,0)=\frac{25}{72} && \left(14313753121 \alpha ^6+31762633248 \alpha ^5+70379308800 \alpha ^4+148842086400 \alpha^3\right.\nonumber\\
&&\left.+273914265600 \alpha ^2+373712486400 \alpha +271790899200\right)e^{-\sigma ^2 \tau ^2} \nonumber\\
  && \times \Bigl[256 \alpha  \left(1352 \alpha ^3+1917 \alpha ^2+2016 \alpha +1152\right)\nonumber\\
 &&+3 \left(1221863 \alpha^4+2584064 \alpha ^3+4755456 \alpha ^2+6488064 \alpha +4718592\right) e^{2 \sigma ^2 \tau ^2} \nonumber\\
   &&-2 \left(1772967 \alpha ^4+3427328 \alpha ^3+5713920 \alpha ^2+7077888 \alpha +4718592\right) e^{\frac{3 \sigma ^2 \tau ^2}{2}}\Bigr]\nonumber\\
   &&\Big/ \Bigl[\left(110286921 \alpha ^5+244372600 \alpha ^4+516812800
   \alpha ^3+951091200 \alpha ^2+1297612800 \alpha +943718400\right) e^{\frac{\sigma ^2 \tau ^2}{2}}\nonumber\\
   &&-4 \left(21489587 \alpha ^5+44324175 \alpha ^4+85683200 \alpha ^3+142848000 \alpha ^2+176947200 \alpha
   +117964800\right)\Bigr]^2. \nonumber
 \end{eqnarray}
   
\begin{eqnarray}
g^{(2)}_{C,p=8}(\tau,0)=\frac{36}{49} && \left(2561459619833 \alpha ^7+5610991223432 \alpha ^6+12450952233216 \alpha ^5+27588689049600 \alpha ^4\right.\nonumber\\
&&\left.+58346097868800 \alpha ^3+107374392115200 \alpha ^2 +146495294668800 \alpha+106542032486400\right)e^{-\sigma ^2 \tau ^2}\nonumber\\
    &&\times \Bigl[25 \alpha  \left(3318119 \alpha ^4+5537792
   \alpha ^3+7852032 \alpha ^2+8257536 \alpha +4718592\right) \nonumber\\ 
   &&\left.-32 \left(21489587 \alpha ^5+44324175 \alpha ^4+85683200 \alpha ^3+142848000 \alpha ^2+176947200 \alpha +117964800\right) e^{\frac{3 \sigma
   ^2 \tau ^2}{2}}\right. \nonumber\\
   && +6 \left(110286921 \alpha ^5+244372600 \alpha ^4+516812800 \alpha ^3+951091200 \alpha
   ^2+1297612800 \alpha +943718400\right) e^{2 \sigma ^2 \tau ^2}\Bigr]\nonumber\\
   &&\Big/ \Bigl[23489061277 \alpha^6+49512008448 \alpha ^5+102122899200 \alpha ^4+197414092800 \alpha ^3\nonumber\\
   &&+329121792000 \alpha ^2+407686348800 \alpha +271790899200\nonumber\\
  && -2\left. \left(14313753121 \alpha ^6+31762633248 \alpha ^5+70379308800 \alpha ^4+148842086400 \alpha^3\right.\right.\nonumber\\
  &&\left.+273914265600 \alpha ^2+373712486400 \alpha +271790899200\right) e^{\frac{\sigma ^2 \tau^2}{2}}\Bigr]^2.\nonumber
\end{eqnarray}

\begin{eqnarray}
g^{(2)}_{C,p=9}(\tau,0)=\frac{49}{2048} && \big(9710015233335279 \alpha ^8+20983477205671936 \alpha ^7+45965240102354944 \alpha
   ^6\nonumber\\
   &&+101998200694505472 \alpha ^5+226006540694323200 \alpha ^4+477971233741209600 \alpha^3\nonumber\\
   &&+879611020207718400 \alpha ^2+1200089453926809600 \alpha +872792330128588800\big)e^{-\sigma ^2 \tau ^2}\nonumber\\
   &&\times \Bigl[36 \alpha  \left(181840923 \alpha ^5+331811900 \alpha ^4+553779200 \alpha ^3+785203200 \alpha ^2+825753600 \alpha +471859200\right)\nonumber\\
   &&+3 \left(14313753121 \alpha ^6+31762633248 \alpha^5+70379308800 \alpha ^4+148842086400 \alpha ^3\right.\nonumber\\
   &&\left.+273914265600 \alpha ^2+373712486400 \alpha+271790899200\right) e^{2 \sigma ^2 \tau ^2} \nonumber\\
   &&-2 \left(23489061277 \alpha ^6+49512008448 \alpha^5+102122899200 \alpha ^4+197414092800 \alpha ^3\right.\nonumber\\
   &&\left.+329121792000 \alpha ^2+407686348800 \alpha+271790899200\right) e^{\frac{3 \sigma ^2 \tau ^2}{2}}\Bigr]\nonumber\\
   &&\Big/ \Bigl[2176979199375 \alpha^7+4603856010292 \alpha ^6+9704353655808 \alpha ^5+20016088243200 \alpha ^4\nonumber\\
   &&+38693162188800 \alpha^3+64507871232000 \alpha ^2+79906524364800 \alpha+53271016243200\nonumber\\
   &&-\left(2561459619833 \alpha ^7+5610991223432 \alpha ^6+12450952233216 \alpha ^5+27588689049600 \alpha ^4\right.\nonumber\\
   &&\left.+58346097868800 \alpha ^3+107374392115200 \alpha ^2+146495294668800 \alpha +106542032486400\right) e^{\frac{\sigma ^2 \tau ^2}{2}}\Bigr]^2.\nonumber
\end{eqnarray}   
   
\begin{eqnarray}
   g^{(2)}_{C,p=10}(\tau,0)=\frac{1024}{81}&& \big(2943285782347428829 \alpha ^9+6292089871201260792 \alpha ^8+13597293229275414528 \alpha^7\nonumber\\
&&+29785475586326003712 \alpha ^6+66094834050039545856 \alpha ^5+146452238369921433600 \alpha^4\nonumber\\
&&+309725359464303820800 \alpha ^3+569987941094601523200 \alpha ^2+777657966144572620800 \alpha\nonumber\\
   &&+565569429923325542400\big) e^{-\sigma ^2 \tau ^2}\nonumber\\
   &&\times \Bigl[49 \alpha  \big(54705318889 \alpha^6+104740371648 \alpha ^5+191123654400 \alpha ^4+318976819200 \alpha ^3 \nonumber\\
   &&+452277043200 \alpha^2+475634073600 \alpha +271790899200\big)\nonumber\\
   &&-8 \big(2176979199375 \alpha ^7+4603856010292 \alpha^6+9704353655808 \alpha ^5+20016088243200 \alpha ^4\nonumber\\
   &&+38693162188800 \alpha ^3+64507871232000 \alpha^2+79906524364800 \alpha +53271016243200\big) e^{\frac{3 \sigma ^2 \tau ^2}{2}}\nonumber\\
   &&+6\big(2561459619833 \alpha ^7+5610991223432 \alpha ^6+12450952233216 \alpha ^5+27588689049600 \alpha^4\nonumber\\
   &&+58346097868800 \alpha ^3+107374392115200 \alpha ^2+146495294668800 \alpha +106542032486400\big) e^{2 \sigma ^2 \tau ^2}\Bigr]\nonumber\\
   &&\Big/ \Bigl[16913362714229743 \alpha ^8+35667627202560000 \alpha^7+75429576872624128 \alpha ^6\nonumber\\
   &&+158996130296758272 \alpha ^5+327943589776588800 \alpha^4+633948769301299200 \alpha ^3\nonumber\\
   &&+1056896962265088000 \alpha ^2+1309188495192883200 \alpha +872792330128588800\nonumber\\
   &&-2 \big(9710015233335279 \alpha^8+20983477205671936 \alpha ^7+45965240102354944 \alpha ^6 \nonumber\\
   &&+101998200694505472 \alpha^5+226006540694323200 \alpha ^4+477971233741209600 \alpha ^3\nonumber\\
   &&+879611020207718400 \alpha^2+1200089453926809600 \alpha+872792330128588800\big) e^{\frac{\sigma ^2 \tau^2}{2}}\Bigr]^2.\nonumber
\end{eqnarray}
For the $\{|A\rangle_c,|D\rangle_d\}$ projection, the results are similar to the ones shown above with some signs flipped so that the `dip' becomes 'peak'.

\end{document}